\documentclass[review, 3p, times]{elsarticle}

\usepackage{amssymb}
\usepackage{relsize}
\usepackage[utf8]{inputenc}
\usepackage{lipsum}
\usepackage{graphicx}
\usepackage{color}
\usepackage{amsmath}
\usepackage{amsfonts}
\usepackage{xcolor}
\usepackage{bm}
\usepackage{epsfig}
\usepackage{float}
\usepackage{soul}
\usepackage[normalem]{ulem}
\sethlcolor{green}
\usepackage{framed}
\usepackage{float}
\usepackage{parskip}
\usepackage[compat=1.1.0]{tikz-feynman}
\makeatother
\usepackage{hyperref}
\hypersetup{
    colorlinks=true,
    linkcolor=green,
    citecolor=blue,
    filecolor=magenta,      
    urlcolor=cyan,
}

\usepackage{hyperref}
\hypersetup{
    colorlinks=true,
    linkcolor=blue,
    filecolor=magenta,      
    urlcolor=cyan,
}

\begin{document}

\begin{frontmatter}

\title{The alloying of first-principles calculations with quasiparticle methodologies for the converged solution of the quantum many-electron states in the correlated compound Iron monoxide}

\author[first,second]{Suvadip Das}
\affiliation[first]{organization={Department of Physics, Birla Institute of Technology and Science Pilani Hyderabad}, city={Hyderabad}, postcode={500078}, state={Telangana},
country={India}}
\affiliation[second]{organization={Department of Physics, George Mason University}, city={Fairfax}, postcode={22030}, state={Virginia},
country={USA}}
\date{\today}

\begin{abstract}
Transition metal oxides belong to a genre of quantum materials essential for the exploration of theoretical methods for quantifying electronic correlation. Finding an efficient and accurate first-principles method for the assertion of such physical properties is momentous for the predictive modeling of physics based thermoelectric and photovoltaic devices. Prior investigations over the last two decades have suggested that incorporation of the so-called random phase approximation (RPA) for the electronic screening interaction by adding up the electron-hole pairs leads to significant improvement in the accuracy of first-principle calculations. Nonetheless, the method has seldom been adapted systematically for studying the properties of prototypical transition metal oxides, particularly that of the correlated d-electronic compound Iron monoxide. In this work, we provide a detailed benchmarking study of a variety of first principles methods such as the density functional theory artificially stabilized by Coulomb interactions, short and long range-separated Hybrid functionals as well as the quasiparticle Green's function approach to self-energy interactions. A rigorous convergence of the self-consistent Dyson's equations have been provided addressing the importance of initial choice of wavefunctions guided by first principles on the converged solutions and the interplay of various orbital degrees of freedom adjacent to the Fermi level. It is momentous to obtain accurate wavefunctions and many-electronic energy states for the quantification of correlation and efficient modeling of oxide interfaces for quantum applications. The study establishes the hybrid functional scheme as the optimal approach for the ideal trade-off between accuracy of the ground state wavefunctions and computational efficiency for large-scale simulations towards the efficient convergence of correlated electronic wavefunctions and low energy electronic properties. 
\end{abstract}


\end{frontmatter}




\section{Introduction}
\label{introduction}
With the advent of an era of quantum technologies and exascale computing, the necessity for large- scale simulations of electronic and phonon interactions in correlated oxides, two dimensional materials and their interfaces in an accurate and cost-effective manner has proven indispensable. The primary objectives of these investigations constitute gaining deeper insights into the rich plethora of quantum phenomenon such as exotic magnetic phases, signatures of geometrical curvatures \citep{Chang_2008}, high temperature superconductors \citep{Warren}, proximity effects in heterostructures {\citep{proximity}}, and multiferroics \citep{PhysRevLett.116.206803} for quantum applications. As a result, the search for a computationally efficient and robust methodology applicable to a variety of prototypical oxides has gained impetus. The scientific quest was further accelerated by recent efforts for transfer learning of interface properties aided by the construction of neural network informed force fields \citep{PhysRevLett.98.146401} and atomistic potentials \citep{Behler}. Numerous experimental discoveries of quantum properties in oxides and their interfaces over the last two decades \citep{Mannhart, Cen, Millis} have laid the framework for practical efforts toward a revolution in quantum devices. These experimental breakthroughs have motivated the exploration of a variety of methodologies for studying materials properties at different length and time scales ranging from density functional theory \citep{Zimmermann, Bechstedt-New}, and ab-initio molecular dynamics \citep{Kresse4, Kresse5} to dynamical mean field theory {\citep{NiO-DMFT3, All4-DMFT1, NiO-DMFT2, NiO-DMFT1,CoO-DMFT1,MnO-DMFT2,MnO-DMFT1,LDA+DMFT}}, and quantum monte carlo simulations \citep{PhysRevB.88.245117}. Among these methodologies, first principles density functional theory, armed with a plethora of advanced exchange correlation functionals have been adopted for most scientific studies on transition-metal oxides till date primarily due to its popularity and success in predicting materials properties.

Oxides of most transition metals are found in abundant quantities in the earth's surface. Research initiatives particularly involving oxides of iron have gained impetus due to their resilience and stability as thin-film substrate materials for photovoltaics and heterostructures, alongside their efficient performance in energy applications such as photo anodes, photo chemistry \citep{YATES20091605} and heterogeneous catalysis \citep{JOZWIAK200717}. Recent success in experimental efforts towards the exfoliation of 2D layered oxides of Iron and related magnetic oxides has reinforced the need to investigate the properties of the oxides in a systematic fashion. \\
\section{Background Landscape}
The history of computational estimation of the structural and magnetic configuration of prototypical oxides dates back several decades. The primary motivation behind the exploration of transition-metal oxides as prototypical test-beds for the development and benchmarking of quantum methodologies, soon transitioned to the exploration of rich quantum phenomenon such as Mott and charge transfer insulators. A highly debated topic in the field of correlated transition-metal oxides is ascertaining whether such a Mott insulator phase exists in this class of materials and understanding the physics behind such a phenomenon. 
Most of the three-dimensional transition metal oxides occur in the rocksalt structure. Amongst them, TiO, VO, RuO${_2}$ and V$_{2}$O$_{5}$ \citep{Khomskii_2014} have been found to be metallic, whereas transition metal oxides such as NiO, CoO, MnO, and FeO \citep{Rodl-G0W0+HSE, FeO-orbitals-DMFT} have been observed to manifest insulating properties with an antiferromagnetic (AFM) ordering \citep{Das} below the Neel's temperature. In the paramagnetic phase, the transition metal (TM) oxides \citep{Das} NiO, CoO, MnO, and FeO have been observed to crystallize in the rocksalt structure, while they undergo magnetic transitions to the antiferromagnetic phase AFM II below the Neel's temperature. The magnetic AFM II phase corresponds to the stacking of simultaneous ferromagnetic layers with alternating up and down magnetic moments along the cubic [111] axis of the crystal lattice and is stabilized by a superexchange interaction mediated by the oxygen atoms.  Interestingly, the magnetic transition is accompanied by a structural transition in most of these transition metal oxides resulting in a reduction of the crystal symmetry brought about either by a rhombohedral, tetrahedral, or orthorhombic distortion. Although several structures \citep{PhysRevLett.108.026403, PhysRevB.86.115134} have been proposed to be associated with the magnetic transition over the years, the exact nature of the structural transition in the iron monoxide compound has been a matter of debate to date. 

Iron oxide, an exemplary transition metal oxide with potential applications, exhibits a multitude of exotic phenomena arising from the interplay of charge, orbital, and spin degrees of freedom. Prevalent in the rocksalt structure, the late transition metal oxide Iron oxide was primarily thought to be a Mott insulator \citep{Terakura, Oguchi} due to the strong correlation effects brought about by the presence of localized d electrons. Later on, further detailed studies revealed the material to be a conventional band-gap oxide, dominated by long range magnetic interactions due to exchange interaction and crystal field effects \citep{PhysRevB.86.115134}. 
Similarly to the other transition metal oxides in this genre, the magnetic phase transitions have been associated with a structural transition leading to symmetry-reduced structures in this correlated oxide. However, unlike the transition metal oxides MnO and NiO, where the associated structural transition reduces the symmetry from rocksalt to that of rhombohedral structure, for FeO, the rhombohedral distortion is accompanied by a strong tetragonal or orthorhombic distortion due to the orbital-order mediated Jahn-Teller effect \citep{FeO-orbitals-DMFT}, thus lowering the symmetry further to that of a monoclinic structure. On the contrary, studies reporting the prevalence of FeO with purely rhombohedral distortions, have rendered the establishment of the exact nature of the structural transition in Iron monoxide difficult. In addition, albeit the crystallographic direction of the magnetic moment lying in the (111) plane for MnO and NiO, several different orientations of the moment have been proposed over time in the case of Iron oxide. 

\begin{figure}
	\centering 
	\includegraphics[width=0.4\textwidth, angle=0]{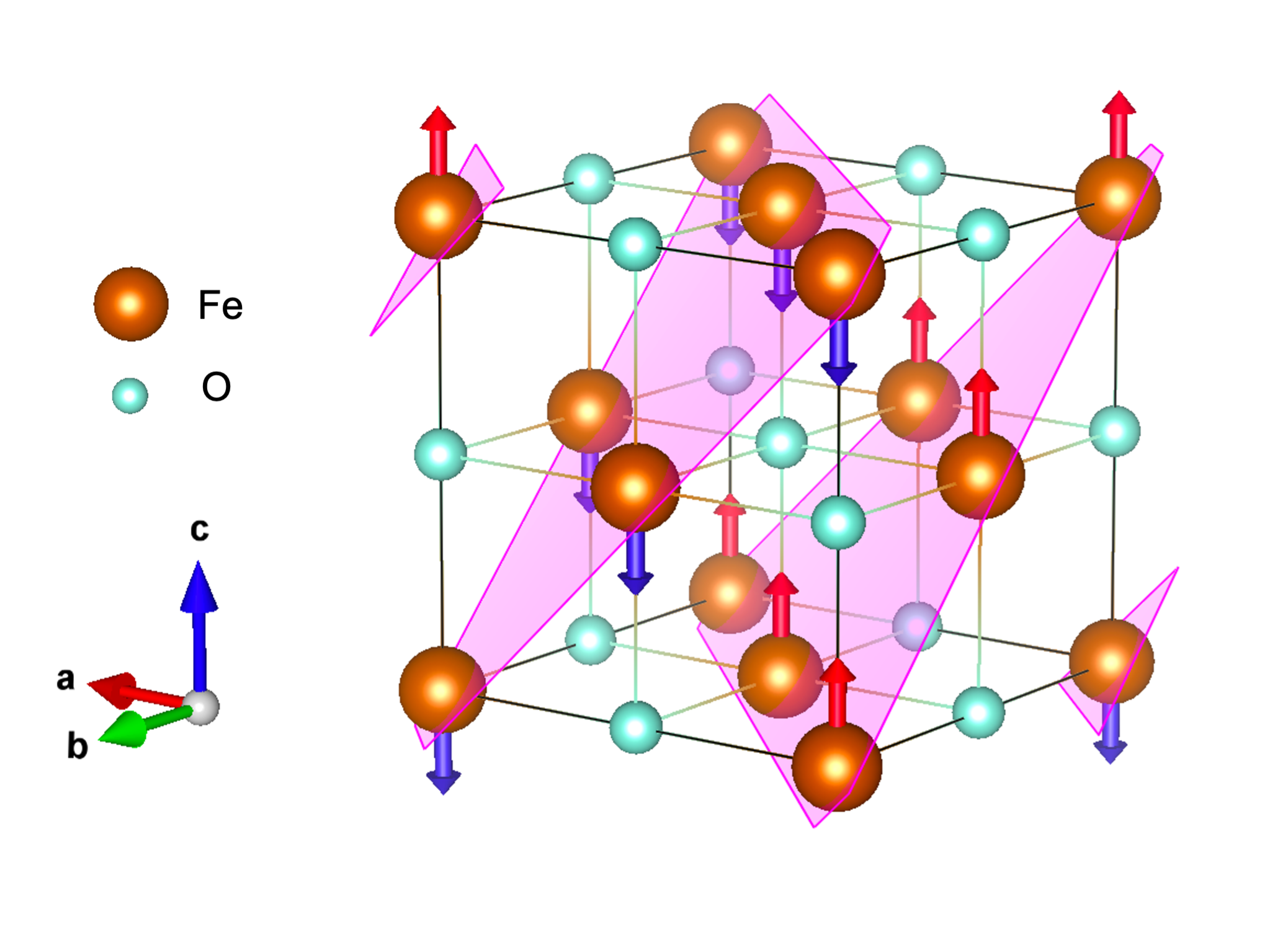}	
	\caption{The crystal structure
    of pristine Iron monoxide displaying the octahedral coordination around the Iron atoms in the rock salt structure. Note that the Iron displays a type II antiferromagnetic (AFM) order where the up and down spins occur in the alternate (111) planes respectively.} 
	\label{fig_0}%
\end{figure}

The inconclusive results detailed above warrant a thorough investigation of the structural phases of Iron monoxide and the search for an accurate yet efficient computational methodology to characterize the material. Density functional theory (DFT) is by far the most promising theory applied for the investigation of the electronic properties of transition metal oxides \citep{Sakuma} till date. Evaluation of the structural and magnetic properties of Iron monoxide such as the lattice constants and spin magnetic moments utilizing a modified version of density functional theory, namely spin-polarized density functional theory, where density functionals of the up and down spin manofolds are treated separately, yields reasonable match with certain experimental findings. Further, computational investigations to understand the structural transitions and orbital ordering utilizing density functional theory \citep{Liechtenstein} reveals the structural distortion in FeO is due to the electron in the $t_{2g}$ subshell of the octahedral crystal field splitting, particularly in the eigenstate of the combination of the correlated transition metal 3d orbitals $\frac{1}{\sqrt{2}}(\left | d_{xz} \right> - \left | d_{yz} \right>)$. Experimental techniques such as the X-ray Photoemission spectra (XPS) \citep{XPS1-Transition-metal-oxides, XPS2-Transition-metal-oxides} and Bremsstrahlung isochromat spectroscopy (BIS) \citep{BIS-Transition-metal-oxides}, also known as inverse photoelectron spectroscopy (IPES) result in electronic bands which are in better agreement with spin-resolved density functional theory compared to other theoretical methods for the evaluation of material properties.

Interestingly, inspite of the remarkable success in predicting materials properties, preliminary first-principles calculations using density functional theory (DFT) erroneously predicted the compound Iron monoxide to be metallic in it's ground state \citep{TMO-Metallic} as opposed to a gapped semiconductor. A thorough all-electron density function theory calculation utilizing the generalized gradient approximation (GGA) indicated the formation of distinct manifolds of bands from the $s$, $p$ and $d$ orbitals respectively, such that the highly dispersive s band crosses the fermi level from above near the $\Gamma$ point, while the localized d bands intersect the fermi level from below resulting in the metallic behaviour of the bulk material. Apart from the inaccurate prediction of characteristic band gap in this material, the photoemission spectra are only at best partially reconstructed utilizing the semilocal exchange correlation functions. 

The lion's share of the incongruity arises from the inability of density functional theory, or for that matter, of any flavor of mean field theory in providing an accurate description of the correlated d electrons in Iron oxide. Efforts towards alleviating the inefficacy in predicting band-gaps while still preserving essential spectral features of photoemission measurements was provided by the density functional theory artificially stabilized by Hubbard interactions \citep{Hubbard, Anisimov} within the so-called LDA+U methodology. A more sophisticated theory for material property prediction, namely the hybrid functions \citep{HSE, HSE06} involving non-local screened exchange interactions, have been accommodated for accurately modeling the low-energy properties of correlated electronic systems over the last decade. Recently, a more rigorous self-consistent perturbative solution of the Dyson's equation, employed to delineate the effect of interactions in terms of the self energy correction has gained popularity for deciphering unresolved phenomenon as well as concocting accurate many-body wavefunctions. However, a detailed investigation for benchmarking the efficacy of the various levels of theories and their systematic application to correlated oxides is still lacking in the literature.   

In the remainder of the manuscript, the amalgamation of first-principles methods with quasiparticle methodologies for accurately modeling the wavefunctions and associated physical properties of Iron monoxide, have been presented with the goal to find the best tradeoff between quantitative accuracy and computational efficiency of the existing formalisms for the computational exploration of correlated electronic systems. 

\section{Theoretical Framework and Computational Methods}
An effective computational recipe well-equipped to aptly model correlated electronic materials is indispensable for the prediction of the band gap in iron monoxide. The search for an accurate methodology is motivated primarily by three principle objectives: (a) Alleviating the band-gap problem, keeping in mind the fact that adiabatic evolution of quantum states does not allow perturbative convergence to a gapped system starting from a metallic ground state. (b) Incorporation of correlated electronic states appropriately to predict orbital ordering of electronic states, magnetic moments and other relevent electronic properties. (c) A quantum method that could potentially describe large scale periodic atomic arrays of oxides and their interfaces, by providing the best tradeoff between computational efficiency and prediction accuracy \citep{Das}.

We discuss three methodologies as viable routes to treating correlated electrons, with the potential to remedy the (wrongly) predicted metallic state using first principles calculations namely (1) Density functional theory (DFT) artificially stabilized by Hubbard interactions, (2) Inclusion of an error-function screened Coulomb interaction in the exchange energy in the so-called Hybrid functionals and (3) Quasiparticle GW approximation \citep{Aryasetiawan, BerkeleyGW} to the self energy for the electron-electron interaction. We will discuss the quality and accuracy of the self consistently iterated solutions of the electronic wavefunctions in the correlated compound Iron monoxide utilizing the above post-DFT formalisms.
\subsection{Density functional theory stabilized by Hubbard Interactions}

The shortcomings of the density functional theory in predicting properties of correlated electronic systems is remedied by the artificial stabilization of the localized electrons by the Hubbard interaction. In this formalism, both the itinerant and localized electrons are treated in a mean field fashion implemented within the density functional approximation, while an additional Hubbard interaction is superimposed on the localized electronic states. The Hubbard interaction \citep{Hubbard} can be delineated as follows:
\begin{equation}
 \textit{H}=\sum_{ij, \, \sigma} t_{ij} \; c_{i, \, \sigma}^{\dagger}c_{j, \, \sigma} + U\sum_{i}^{} n_{i \uparrow} \; n_{i \downarrow} 
\end{equation}
 The first term applies to the itinerant electrons, where the hopping term scales as the band width of the bands corresponding to a particular orbital, whereas the second term signifies the Hubbard interaction \citep{Fuchs08, TMO-Metallic}, responsible for the splitting between upper and lower Hubbard bands.
\begin{equation}
E_{LDA+U}[\,\rho (\mathbf{r})\,] = E_{LDA}[\, \rho (\mathbf{r})\,] + E_{Hub}\,[\,{\{n_{mm^{'}}\}^{l\sigma}}] - E_{dc}\,[\, \{n^{l\sigma}\}] 
\end{equation}

The incorporation of the Hubbard interaction energy in density functional theory is accompanied by the simultaneous removal of the double counting energy error arising from the mean field treatment of the relevant localized electrons. Here the $n, l,$ and $\sigma$ refer to the occupation number, site and angular momentum projection indices of the electrons respectively.
\subsection{Hybrid functionals}
Another popular methodology for the convergence of many-electron wavefunctions in quantum solids pertains to the hybrid functional scheme. This formalism hybrids together the density functional exchange energy with that of the Hartree Fock, thereby appropriately modeling electronic screening and efficiently predicting band-gaps underestimated by density functional theory. Further, the hybrid functional technique drastically reduces the computational efficiency of the wavefunction based Hartree Fock methods and serves as the best of both worlds. Based on the mixing of first principles methods, the hybrid functional technique offer a wide choice of mixing parameters and coulomb interaction cut off parameters.
We will evaluate and bechmarking two popular methods: 1. The B3LYP (Becke, 3-parameter,Lee–Yang–Parr) \citep{B3LYP, B3LYP1} exchange correlation functional, which mixes the Hartree Fock and Vosko-Wilk-Nusair parameterization of the Local Spin Density functional Approximation (LSDA) \citep{PhysRevB.22.3812}, as well as the Becke88 exchange functional for the exchange energy 2. The Heyd, Scuseria and Ernzerhof (HSE) functional \citep{HSE, HSE06} which cleverly utilizes a error function screened Coulomb interaction, such that only the short range part of the exchange energy from Perdew, Burke, and Ernzerhof (PBE) density functional is combined with the short range of the Hartree Fock approximation, while the long range interaction is treated separately.
\begin{equation}
 E_{xc}^{\mu \;HSE} = aE_{xc}^{HF,\; SR}(\mu) + (1-a) E_{xc}^{PBE,\; SR}(\mu) + E_{xc}^{PBE,\; LR}(\mu) + E_{c}^{PBE}
 \end{equation}
Depending on the choice of the parameter $\mu$ as 0.2 or 0.3, and the parameter $a$ to be 0.25, the functionals are defined as Hybrid functional flavours named HSE06 and HSE03 respectively \citep{HSE06}. The drawbacks of the Hybrid functionals for materials modeling includes a logarithmic divergence of orbital energies in the case of metals, and the exponential decay of the exchange interaction with band gap for solids. Inspite of certain shortcomings, the hybrid functionals is, by far, one of the most accurate and popular methodologies for the first principles and many body calculations in crystalline solids. 
\subsection{The Quasiparticle GW Methodology and Random Phase Approximation}

\begin{figure}[ht]
\centering
\[ 
\vcenter{\hbox{\begin{tikzpicture}
  \begin{feynman}
    \vertex (i);
    \vertex [right=1.25cm of i] (o);
    \diagram*{
      (i) --[double,double distance=0.3ex,thick,with arrow=0.5,arrow size=0.2em] (o)    
    };
  \end{feynman}
\end{tikzpicture}}}
~=~
\vcenter{\hbox{\begin{tikzpicture}
  \begin{feynman}
    \vertex (i);
    \vertex [right=1.25cm of i] (o);
    \diagram*{
      (i) --[fermion] (o)     
    };
  \end{feynman}
\end{tikzpicture}}}~+~
\vcenter{\hbox{\begin{tikzpicture}
  \begin{feynman}
    \vertex (i);
    \node[right=1.0cm of i,draw,fill= gray,circle] (v){$\Sigma$};
    \vertex [right=1.25cm of v] (o);
    \diagram*{
          (i) --[fermion] (v),        
      (v) --[double,double distance=0.3ex,thick,with arrow=0.5,arrow size=0.2em]  (o)
    };
  \end{feynman}
\end{tikzpicture}}}
\]
\vspace{-1.0 cm}

\tikzset{graviton/.style={decorate, decoration={snake, amplitude=.4mm, segment length=1.5mm, pre length=.5mm, post length=.5mm}, double}}

\[
\vcenter{\hbox{\begin{tikzpicture}
\begin{feynman}[baseline=(i.base), xshift = -0.5 cm, yshift = 0.25 cm]
    \vertex (i);
    \vertex [right=0.25cm of v] (o);
    \node[right=0.0cm of i,draw,fill= gray,circle] (v){$\Sigma$}; \quad \quad
\end{feynman}
\hspace{0.75 cm} \vspace{0.5 cm} $=$;
\begin{feynman}[baseline=(v.base)]
    \vertex (a);
    \vertex [right=1.25cm of a] (b);
    \vertex [right=1.15cm of b] (c);
    \diagram*{
      (b) --[double,double distance=0.3ex,thin,with arrow=0.5,arrow size=0.2em] (c)    
};
\feynmandiagram[layered layout,inline=(b.base),horizontal=a to c] {
a -- [draw=none, out=135, in=45, min distance=0.5 cm, text=black] b
  -- [graviton, out=135, in=45,  min distance=1.5cm] c -- [charged scalar] c,
};
  \end{feynman}
\end{tikzpicture}}}
\]

\vspace{0.75 cm}
\tikzset{graviton/.style={decorate, decoration={snake, amplitude=.4mm, segment length=1.5mm, pre length=0mm, post length=0mm}, double}}

\tikzset{every picture/.style={
    scale=0.7, transform shape,
    }}
\begin{tikzpicture}
    \begin{feynman}
\vertex (a);
\vertex [above left= of a] (b);
\vertex [right= of a] (c);   
\diagram[layered layout,inline=(a.base),horizontal=a to c] {
(a) -- [graviton, out=0, in=180] (c);
};
\end{feynman}
\end{tikzpicture} \quad \quad
\begin{tikzpicture}[baseline=(a.base)] 
{$=$}
\end{tikzpicture}\quad \quad \\
\begin{tikzpicture}[baseline=(b.base)]
\begin{feynman}[ small]
\hspace{0.5 cm}
\vertex (a);
\vertex [above left= of a] (b);
\vertex [below left= of a] (c);
\vertex[right= of a] (d);
\vertex[right= of d] (e);
\vertex [right= of e] (f);
\diagram {
(a) -- [boson] (d);
(d) -- [fermion, half left] (e);
(e) -- [fermion, half left] (d);
(e) -- [boson] (f);
};
\end{feynman}
\end{tikzpicture}\quad \quad
\begin{tikzpicture}[baseline=(b.base)]
\hspace{0.25 cm} {$+$} \hspace{0.5 cm}
\end{tikzpicture}\quad \quad \quad
\begin{tikzpicture}[baseline=(b.base)]
\begin{feynman}[small]
\vertex (a);
\vertex [above left= of a] (b);
\vertex [below left= of a] (c);
\vertex[right= of a] (d);
\vertex[right= of d] (e);
\vertex [right= of e] (f);
\vertex [right= of f] (g);
\vertex [right= of g] (h);
\diagram {
(a) -- [boson] (d);
(d) -- [fermion, half left] (e);
(e) -- [fermion, half left] (d);
(e) -- [boson] (f);
(f) -- [fermion, half left] (g);
(g) -- [fermion, half left] (f);
(g) -- [boson] (h);
};
\end{feynman}
\end{tikzpicture}\\
\begin{tikzpicture}[baseline=(b.base)]
\hspace{0.5 cm} {$+$} \hspace{0.5 cm}
\end{tikzpicture} \quad \quad \quad \quad
\begin{tikzpicture}[baseline=(b.base)]
\begin{feynman}[small]
\vertex (a);
\vertex [above left= of a] (b);
\vertex [below left= of a] (c);
\vertex[right= of a] (d);
\vertex[right= of d] (e);
\vertex [right= of e] (f);
\vertex [right= of f] (g);
\vertex [right= of g] (h);
\vertex [right= of h] (i);
\vertex [right= of i] (j);
\diagram {
(a) -- [boson] (d);
(d) -- [fermion, half left] (e);
(e) -- [fermion, half left] (d);
(e) -- [boson] (f);
(f) -- [fermion, half left] (g);
(g) -- [fermion, half left] (f);
(g) -- [boson] (h);
(h) -- [fermion, half left] (i);
(i) -- [fermion, half left] (h);
(i) -- [boson] (j);
};
\end{feynman}
\end{tikzpicture} \quad 
\begin{tikzpicture}[baseline=(b.base)]
{$+$};
\end{tikzpicture} \quad \quad
\begin{tikzpicture}[baseline=(b.base)]
{$....$};
\end{tikzpicture}

\caption{The Feynman diagrams explaning the solution of the Dyson's equation, where the self-energy term is approximated by the self-consistent $GW$ method\citep{Hedin, Aryasetiawan, Hedin-1}. Note that the screened coulomb interaction is evaluated by inclusion of the electron-hole bubbles in the polarization upto infinite orders within the Random phase approximation ($RPA$).}
\label{fig_1}%
\end{figure}
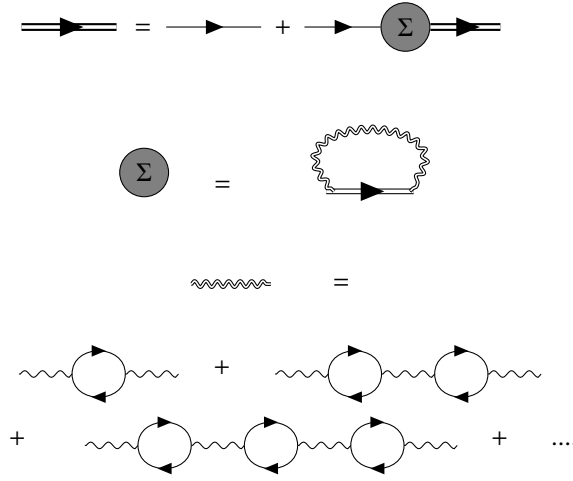

Recent development in the last decade has led to the inclusion of the effects of electron-electron, electron-phonon and electron-magnon interactions in quantum simulations of real materials by the so-called quasiparticle Green's function methodology \citep{Hedin, Hedin-1}. In this many-body approach, interactions are included in the form of a non-local self-energy term $\Sigma(\vec{r},\vec{{r}'}, \epsilon_{nk}^{QP})$, which depends on the quasiparticle energies of the system.
The solution of the many-body wavefunction then amounts to solving the coupled set of equations for the electrons in the system, which can in principle, be solved perturbatively provided that the interaction Hamiltonian is a fraction of the Hamiltonian of the non interacting system.
\begin{equation}
H_{0}^{DFT}(\vec{r})\: \psi _{nk}(\vec{r})+\int d^{3}\vec{r}\;\; \Sigma(\vec{r},\vec{{r}'}, \omega)\: \psi _{nk}(\vec{{r}'}) = \epsilon_{nk}\: \psi _{nk}(\vec{r}) 
\end{equation}

In this equation, the electronic self-energy can be reformulated as a convolution of the full electronic Green's function ($G$), the screened Coulomb interaction ($W$) and the three-point vertex correction ($\Gamma$). The self-consistent iteration of the full Green's function amounts to starting out with the initial choice of Green's function($G^{DFT})$ and iteratively updating the poles of the Green's function until convergence is achieved ($G^{QP})$ and the self-consistent solution of the Dyson's equation has been acquired.

The conventional solution to the converged Green's function, and hence density can be obtained by self-consistent solution of the so-called Dyson's equation \citep{Hedin, Aryasetiawan}:
\begin{equation}
G = G_{0}\, + \,G_{0}\,\Sigma\,G
\end{equation}
The self-energy due to interactions can be expanded in iterative orders of the convolution of the Green's function and screened Coulomb interaction as shown below:
 \citep{Hedin, Aryasetiawan}
\begin{align}
\Sigma &=  i\, G(1,2)\, W(1^{'},2^{'})\, \Gamma(1,{1}';2,{2}')\nonumber \\
       &=  i\, G(1,2)\, W(1,2)\, - \int d^{3}\int d^{4}\: G(1,3)\, G(3,4)\, \times\nonumber\\
       & \hspace{6mm}G(4,2)\, G(1,4)\, W(3,2)\, +\: .....
\end{align}
For the sake of computational resource constraints, it is customary to consider the first term in the expansion, the correction of the higher orders contributing to a small fraction, at best. This method is referred to as the popular GW Approximation to the Self-energy. Focusing on the spectral signatures thus obtained, the position of the quasiparticle peaks can be identified to be located at $E_{i}=\epsilon^{DFT}_{i}-Re\; \Delta\Sigma _{i}(\omega)$ and the lifetime of decay of the quasiparticle peaks due to interaction is given by $\frac{1}{Im \; \Sigma _{i}(\omega)}$. The renormalization factor, which determines the amplitude of the quasiparticle peaks, is expressed as 
\begin{equation}
 Z_{i}=[1-\frac{\partial\: Re\: \Delta \Sigma_{i}
(E_{i})}{\partial \omega}]^{-1}  
\end{equation}
and is an indicator of the localization of the orbital-resolved quasiparticle peaks.
Next, we elucidate on the recipe for treating screened Coulomb interaction in our computational methodology. We apply the so-called random phase approximation for modeling the screening of the electron-electron interaction particularly efficient for systems with high electronic densities. In this formalism, the bubble diagrams for the electron-hole pairs are added up recursively in the polarization term (P) of the system with strong electron-electron interaction. The screened Coulomb interaction can then be written as a geometric series of electron-hole pairs \citep{10.1063/1.3250347} added up to infinite orders:
\begin{equation}
W_{RPA}= \frac{(4\pi e^{2}/q^{2})}{1+(4\pi e^{2}/q^{2}) P}
\end{equation}
Figure 2 elucidates the Feynmann diagrams for the computational recipe of the quasiparticle GW methodology. The first row represents the solution of the Dyson's equation, while the second row delineates the expansion of interacting self-energy term as a convolution of the electronic Green's function and the screened Coulomb interaction potential. The third row reflects the expansion of the various orders of the popular random phase approximation in terms of the electron-hole pairs (or bubble diagrams) as shown in the Feynman diagrams in Fig. 2.

\section{Results}
\subsection{Computational Details}
In this section, we provide the essential details of computational techniques based on first-principles calculations and many-body perturbation theory for the determination of the correlated electronic states in iron monoxide. We have employed the Perdew-Burke-Ernzerhof (PBE) \citep{PBE} functional form of the Generalized Gradient Approximation (GGA) exchange correlation functional \citep{David-Singh-1} and the plane-wave norm-conserving pseudopotentials as implemented in the Vienna Ab Initio Simulation Package (VASP) \citep{VASP}. Our DFT+U calculations  involve the Generalized Gradient Approximation (GGA) exchange correlation functional artificially stabilized by the Hubbard interaction. We resort to the Dudarev approach \citep{Dudarev} where a rotationally invariant formulation of DFT+U has been implemented, such that instead of treating the Hubbard interaction U and Hund's correction J individually, an effective parameter $U_{eff} = U - J$ has been incorporated in our calculations. The pseudopotentials for Iron (Fe) consisted of 3d and 4s electrons in the valence state while that of oxygen (O) contained 2s and 2p electrons in the valence states respectively. The single particle wavefunctions were evaluated within DFT with a plane-wave energy cutoff of 600 Ry. 

The crystal structure used \citep{FeOlattice} in all our calculations is the rocksalt cubic wustite structure (space group $\mathit{Fm\bar{3}m}$) with lattice parameter $a = 8.666 \AA$ in conformity with the International Crystal Structure Database
(ICSD) and Materials Project Repository \citep{Jain2013} . It consists of Fe and O atoms arranged to form a face centered cubic lattice, with the
interpenetrating lattices forming a 3D checkerboard like pattern as shown in Fig. 1. Each Fe (and O) atoms occupy the vertices of the octahedra, surrounded by 6 octahedrally coordinated O (and Fe) atoms in the form of interpenetrating octahedrons thus creating the networks of 3D checkerboard pattern. The relaxed in-plane and out of plane lattice constants of $a =$ 8.67$\AA$ for bulk
FeO in our calculations exhibit slight mismatch with the experimental lattice
parameters {\citep{FeOlattice}} . Small distortions from the cubic structure have been ignored in the benchmarking of the quasiparticle
GW calculations \citep{Kotani} to accomodate computational resource constraints. 

The polarizability matrices and quasiparticle energies were evaluated using Brillouin zone sampling grids \citep{MP76} of $4\times4\times4$ for the priodic crystalline lattices.
The choice of the size of k-point sampling grid is acceptable for similar studies, and it was noted that the convergence of the quasiparticle energies exhibits at best, a weak dependence on the size of the k-point grid. A maximum of 144 and 88 occupied bands were utilized for the Density functional theory and quasiparticle GW calculations respectively. We further note that, the perturbative solution of the Dyson's equation was performed in the GW approximation utilizing a k-point grid of $4\times4\times4$, a maximum of 88 bands and 64 frequency ($\omega$) points for the evaluation of the frequency dependence of the response functions within the random phase approximation.

\subsection{Implementation of the GW Methodology}

There exists a multitude of variations and flavors of the implementation of the GW methodology for studying material properties in literature. Here we elucidate the particular form of implementation utilized in our calculations for benchmarking of properties in Iron monoxide. In this calculation, a variant similar to that proposed by Schilfgaarde et. al. 
\citep{Schilfgaarde, Shishkin1} has been utilized. Our methodology relies on the explicit evaluation of frequency-dependent response functions, and is different from the incorporation of the plasmon-pole model or the implementation
without the explicit calculation of virtual orbitals popularly used to reduce computational requirements. Some of the popular
flavors of the GW method include the one-shot quasiparticle GW calculation also known as $G_{0}W_{0}$ , the partially
self-consistent GW method ($EVGW_{0}$ ), the partially self-consistent quasiparticle GW method $QPGW_{0}$ , the full quasiparticle self-consistent GW (QPscGW ) and the low
scaling, fully self-consistent quasi-particle GW calculations (QPGWR).

In an effort to perform our qusiparticle self-consistent GW (QPscGW) calculations \citep{Hybertsen, VaspGW1, VaspGW2}, we choose a semi-local exchange correlation potential within the GGA+U approximation as the initial starting point and solve the GW equations iteratively \citep{Shishkin3, Fuchs08}. 
The exchange correlation potential is updated at every step of iteration $i$ by 
linearizing the self-energy $\Sigma^{(i-1)}(\epsilon)$ obtained in the previous iteration $(i-1)$
around the quasiparticle energy eigenvalues $\epsilon^{(i-1)}_n$ 
obtained in the former step. Thus, we solve the following Hamiltonian equation: 
\begin{equation}
\textbf{H}^{(i)}\mid \psi^{(i)}_{n}\rangle = \epsilon^{(i)}_n \textbf{O}^{(i)} | \psi^{(i)}_{n}\rangle,
\end{equation}
where the Hamiltonian and overlap matrices are given by:
\begin{eqnarray}
\textbf{H}^{(i)}&=&T\; +\; V_{ext}\; +\; V_{H}\; +\; V_{xc}^{(i)}, 
\label{hamiltonian}\\ 
V_{xc}^{(i)} &=& [\Sigma^{(i-1)}(\epsilon)\; -\; \epsilon \frac{\partial \Sigma^{(i-1)}(\epsilon) }{\partial {\epsilon}}]|_{\epsilon_{n}^{(i-1)}}, \label{exc-corr} \\
\textbf{O}^{(i)} &=& [1 -\; \frac{\partial \Sigma^{(i-1)}(\epsilon)}
{\partial \epsilon}|_{\epsilon_{n}^{(i-1)}} ].
\end{eqnarray}
The Hamiltonian operator $\textbf{H}^{(i)}$ and the overlap operator $\textbf{O}^{(i)}$ can be expressed in an appropriate basis set
$|\phi_n \rangle$, and we take into account only the Hermitian part of the self-energy and overlap matrix
in this basis \citep{Shishkin3}.
As a consequence, we finally solve the corresponding Hermitian eigenvalue problem
\begin{eqnarray}
\tilde U^{\dagger} \tilde O^{-1/2} \tilde H \tilde O^{-1/2}  \tilde U = \Lambda,
\end{eqnarray}
where $\tilde U$ is a unitary matrix and $\Lambda$ is the diagonal eigenvalue
matrix, in terms of which the updated wavefunctions are given as
\begin{eqnarray}
| \psi^{(i)}_n \rangle = \sum_m \tilde U_{nm} | \phi_m \rangle.
\end{eqnarray}

In the next few sections, we elucidate on the implications of the above formalism for the transition-metal oxide, Iron monoxide, and discuss the merits and drawbacks of utilizing the abovementioned recipe for evaluating materials properties. 

\subsection{Adiabatic evolution of eigenstates and orbital ordering}
\begin{figure*}[h!]
	\centering 
	\includegraphics[width=0.9\textwidth, angle=0]{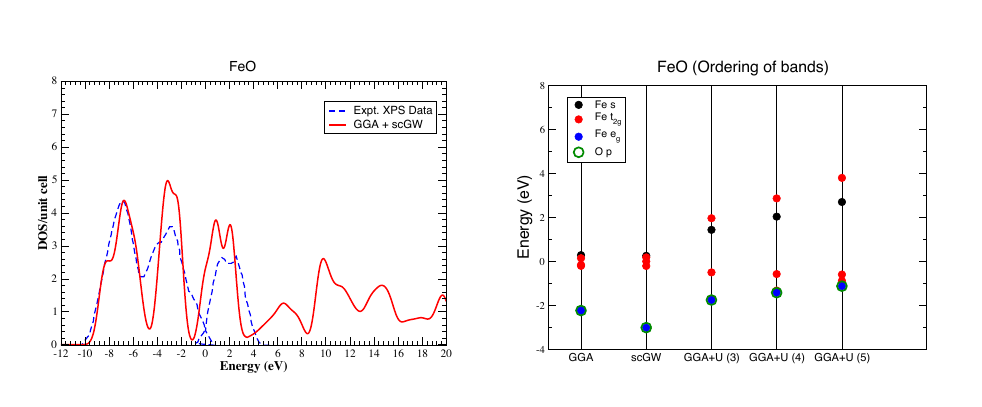}	
	\caption{
(a) The density of states for electronic states as obtained from GGA calculations are compared to the experimental XPS Data for occupied energy levels for FeO near the Fermi level. Note that GGA fails to reproduce the energy band gap obtained in this material. (b) A closer look at the orbital ordering of the bands near the fermi level shows that GGA starts with a different ordering in bands, thus leading to discrepancy with the measured band gaps. It is crucial to include the Hubbard-like
U interactions on the localized d electrons of Iron in order to correctly predict the energy spectrum in these oxides.
    } 
	\label{fig_2}%
\end{figure*}
Prediction of accurate quasiparticle band gaps and magnetic moments in transition-metal oxides have been a long standing challenge in theoretical and computational quantum materials physics. It is crucial to benchmark various levels of theory in order to accurately model transition-metal oxides, which act as potential test-beds for application of novel approaches and theoretical models. In this connection, the quasiparticle GW methodology has proven to be effective in providing accurate low-energy properties such as band gaps, density of occupied states, formation energies and magnetic moments. 
During the benchmarking of various levels of theory, we have taken utmost care so that our calculations are aimed at addressing three major facts: (1) the accuracy of our theoretical predictions in matching the results from experimental findings (2) the reproducibility of the correct orbital ordering of the energy levels (3) computational efficiency of the convergence of the quasiparticle wavefunctions by the iterative solution of the parameter-free Dyson's equations.

During first-principles and many body calculations of transition metal oxides, the incidence of energy level crossings bears remarkable implications on the lattice, spin and orbital degrees of freedom governing the physics in this class of materials.
When such level crossings occur during the course of convergence of the wavefunctions in the iterative QPscGW scheme for solving the Dyson's equation perturbatively, the adiabatic nature of the evolution of quantum eigenstates cannot be guaranteed. According to the adiabatic theorem of quantum mechanics, during the time evolution of a quantum state, the eigenstates obtained by solving the Schrodinger’s equation must be in one-to-one correspondence with the initial eigenstates. To maintain conformity with the above, and to be in correspondence with the perturbed Hamiltonian, level crossings of the electronic bands should not be accounted for, during the process of the time evolution. 
Hence, in order to ensure adiabatic evolution of the electronic states, it is crucial that the orbital ordering of the energy states prior to applying our iterative QPscGW scheme matches with that of the final energy states. The above justifies a validity constraint on our starting eigenstates from various levels of theoretical approximations for the prototypical rock-salt transition metal oxide Iron monoxide. With these constraints, we have evaluated in detail the convergence of the many-body ground state for the transition-metal oxide.

\begin{figure*}[h!]
	\centering 
	\includegraphics[width=0.9\textwidth, angle=0]{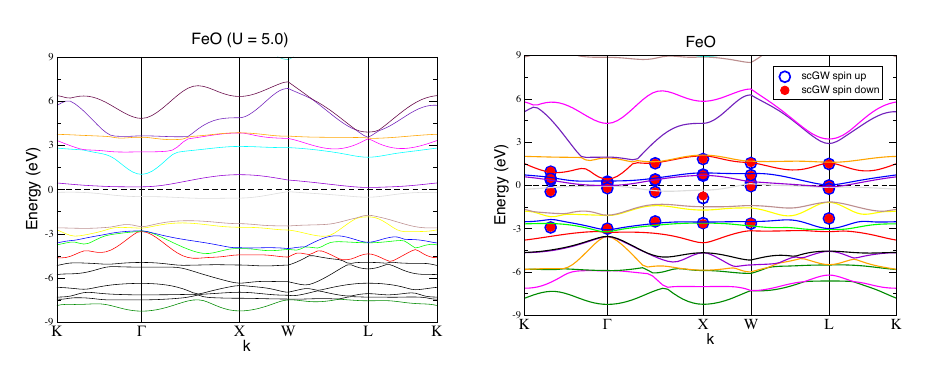}	
	\caption{
The energy bands of FeO near the Fermi level as obtained from GGA+U calculation with (a) U = 5 eV and (b) self-consistent GW by solving the Dyson’s equation iteratively on top of GGA+U. The GW eigenvalues for up and down spin are shown in blue and red respectively.
    } 
	\label{fig_3}%
\end{figure*}

We performed our first-principles calculations utilizing the Generalized Gradient approximation of the spin-resolved density functional theory to evaluate the electronic structure of the transition metal oxide. Figure 3 (a) displays a comparison of the experimental photoemission (XPS) spectra and Bremsstrahlung isochromat spectroscopy (BIS) \citep{XPS1-Transition-metal-oxides,XPS2-Transition-metal-oxides, BIS-Transition-metal-oxides} for the occupied states (shown in dotted blue) for Iron oxide and the corresponding plots for density of localized states as obtained by iteratively solving the GW quasiparticle wavefunctions starting from the Generalized Gradient approximation. Note that in the case of Iron oxide, the spin-resolved GGA calculations fail to reproduce the non zero electronic band gap due to strong electronic correlations in this material. We note that while density functional theory accurately predicts the band gaps in $\mathit{MnO}$ and $\mathit{NiO}$, it does not succed in predicting the band gap in the other correlated material $\mathit{CoO}$. In order to include the role of correlation in these materials, one has to include the effect of Hubbard U term and utilize advanced electronic structure methods such as the popular hybrid functional calculations. However, interestingly, even though DFT predicts the material to be metallic, the spectral signature of the density of states match agreeably with the experimental XPS peaks \citep{Das, TMO-Metallic}. 

The atom and orbital decomposed partial density of states of the bands for various levels of theory are displayed in Fig. 3(b). We
note that the GGA approximation exhibits Fe $t_{2g}$ states
at the valence band maxima as well as the conduction band minima. However, due to the repulsion between the upper and lower Hubbard bands, these $t_{2g}$ levels are
pushed away exposing the Fe s state at the conduction band minimum. As a result of the above, application
of the perturbative quasiparticle GW correction \citep{Das} to the GGA wavefunction does not alter the ordering of bands predicting the material to be a metal. However, the starting wavefunctions obtained from GGA+U opens up an electronic
band gap in the material and appropriately describing
Iron oxide to be an antiferromagnetic insulator. It is evident that our oxide material would be described by the
Hubbard model only in a regime of values of U associated with a specific orbital ordering in the material. An alteration in the orbital ordering is forbidden as suggested by the adiabatic theorem of quantum mechanics.

\subsection{Alleviating the band gap problem with Hubbard interaction}

The fact that the spectral signatures of the experimental XPS spectra agree reasonably well with that of the GGA approximation hint towards the fact that the physics of itinerant electrons are captured by our first principles calculations. Next, we address the addition of the Hubbard interaction in order to accurately treat correlation effects due to the localized electronic states. It has been predicted that an Hubbard U value of (0.6 - 1.0) eV yields GGA+U results that match well with experimental band gaps \citep{Choice-U1} in the case of the correlated compound $FeO$.
However, recently a much needed formalism for applying the linear response approach to obtain a quantitative estimation of the effective interaction parameter U in the LDA+U approximation has been implemented. Application of these schemes have provided suggested U values in the range (3 - 7) eV in the case of the transition metal oxide $FeO$ \citep{Choice-U2, Noa}. As mentioned in the previous section, the ordering of the orbital projected band structure reveals that the orbital ordering of the bands remain unchanged for U values between (3 - 6) eV and agrees with the experimental findings \citep{Choice-U2, Das, Noa}. First principles calculations utilizing a bayesian optimization assisted formalism for determination of the U values reveal much higher values of the Hubbard interaction for $FeO$. Fig. 4(a) shows the electronic structure obtained by solving the GGA+U equations with Hubbard parameter U = 4.0 eV along certain special k-point paths within the Brillouin zone. We note that the GGA+U method alleviates the band-gap problem and results in a band gap with the conduction and valence edges located at the commensurate k-points K and $\Gamma$ respectively. 

Fig. 4(b) displays the results of self-consistent solution of the quasiparticle GW equations starting from the GGA+U wavefunctions obtained from the previous step. The GW eigenvalues for up and down spin are shown using blue and red colors respectively. We observed a highly dispersive $s$ band located at higher energies and is more pronounced at the $\Gamma$ point, while the localized $d$ bands result in narrow bands near the Fermi level. The occupied bands near the valence band edges are prominently due to hybridization of the $p$ bands with that of the $d$ bands. Interestingly, the deviation from the GGA+U bands is more pronounced near the commensurate k-point $X$ due to renormalized screening effects included in the $GW$ formalism.

\begin{figure*}
	\centering 
	\includegraphics[width=0.9\textwidth, angle=0]{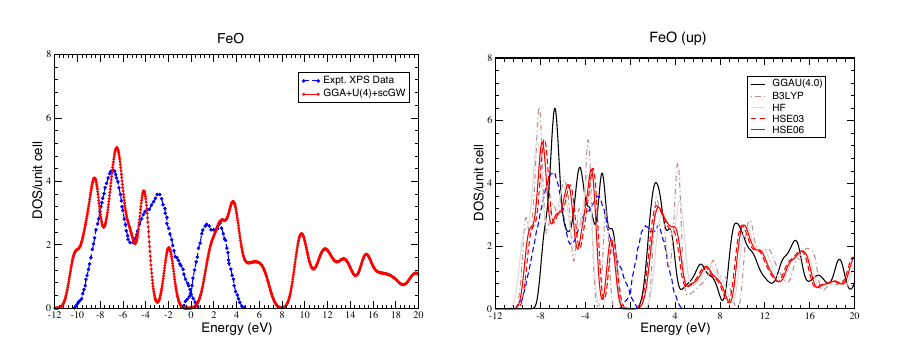}	
	\caption{
(a) The converged density of states for up spin electrons as obtained from applying GGAU + self-consistent GW corrections. Note that the GW quasiparticle structure captures the important features of the valence and conduction electron
states. (b) A comparison of density of states from various levels of theory from Hartree Fock and B3LYP to Hybrid functionals (HSE03 and HSE06). The Hybrid functionals match closely with the positions and amplitudes of the experimental spectra and serves as the best trade-off between accuracy and computational cost for the oxide of Iron.
    } 
	\label{fig_4}%
\end{figure*}

A comparison of the density of states of the converged GW calculations starting with GGA+U are shown in Fig. 5(a). We observed that the convergence of the band gap as a function of the iterations changes slowly and monotonically, until convergence was achieved. Though it is common to apply a single shot $G_{0}W_{0}$ calculation starting from the GGA+U calculations, they are not, by any means, sufficient to yield accurate band gaps in the case of Iron monoxide. 

The convergence of the Dyson's equation being a self-consistent iterative process, the choice of the initial wavefunctions are crucial for efficient convergence of the GW methodology. In Fig. 5(b), we exhibits results for the evaluation of the density of states from various levels of theory. The figure exhibits a comparison of the spectral signature of the electronic states obtained from GGA+U (U = 4.0 eV), B3LYP functional, the Hartree Fock method, and the popular Hybrid functional (HSE03 and HSE06) method respectively \citep{Coulter-scGW, Rubio-HSEtest}. Interestingly, all the levels of theory provide spectral decomposition of the electronic states which agree well with the experimental XPS and BIS results, particularly the four prominent peaks distinctly visible in the occupied density of states and their band widths obtained from our calculations. Specifically, the calculations using the hybrid functional approaches (HSE03 and HSE06) result in electronic states which replicates the band edges as well as the spectral width of the states within the predicted energy window. The spectral width of the occupied density of states for the Hartree Fock calculations are spread out over a broader energy window, which indicates the possibility of larger time to convergence of the Dyson's equation. The GGA+U results agree well with the experimental band edges, spectral widths of occupied states as well as the amplitudes of the density of states. The density of states using the B3LYP functional \citep{B3LYP} results in larger peaks, which significantly overestimate the amplitudes of the density of states. Our calculations indicate GGA+U and Hybrid functionals \citep{HSE, HSE06} (HSE03, HSE06) as potentially the best starting wavefunctions for the convergence of the self-consistent GW calculations. 

\subsection{Choice of starting wavefunctions from DFT+U and HSE for convergence of QPscGW band gaps}
\begin{figure*}[ht!]
	\centering 
	\includegraphics[width=0.9\textwidth, angle=0]{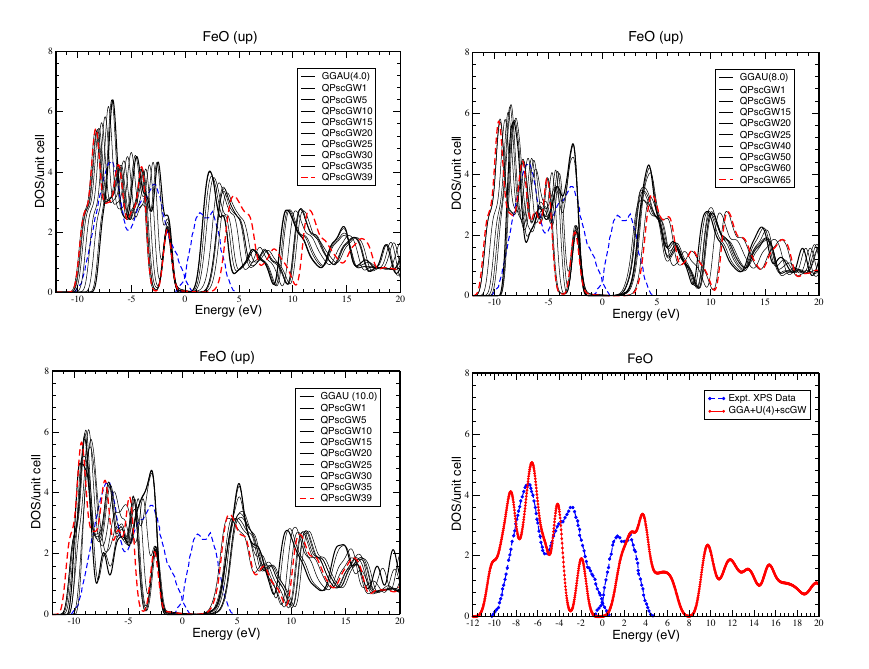}	
	\caption{
The convergence of the electronic wavefunction as a function of GW iterations starting from (a) U = 4.0 eV, (b) U = 8.0 eV and (c) U = 10.0 eV. We note that the converged results starting from U = 8.0 and 10.0 eV shows overestimated band
gaps when compared to experimental spectra. (d) The converged quasiparticle results starting from GGA+U with U = 4.0 eV
provides good quality of converged wavefunctions at a cost of higher computational effort.
    } 
	\label{fig_5}%
\end{figure*}

In this section, we provide a detailed comparison of the converged wavefunctions starting from various levels of theory. Let us consider the fully converged solution of the single-particle Green's function (say $\tilde{G}$), which is obtained from the Dyson's equation, and is essentially a fixed point solution of an iterative scheme. Thus, the starting Green's function for the iterative process is $G_{0}$, while the nth and (n+1)th iterative solutions are given by $G_{n}$ and $G_{n+1}$ respectively. Since the different choice of starting Green's functions are connected to the fixed point solution through analytical continuation, the final solution $\tilde{G}$ should, in principle, be independent of the starting wavefunction $G_{0}$.
Hence, in an effort to make the quasiparticle GW method computationally efficient, we search for an initial wavefunction that is closer to the converged solution, so that, faster convergence can be achieved \cite{Hedin,Schilfgaarde}. 

In this connection, we address the following concerns:
(1) Whether we require too many iterations of the quasiparticle self-consistent GW (QPscGW) method \citep{Schilfgaarde} for practical applications to real materials. 
(2) The dependence of the convergence rate on the initial choice of the wavefunctions, particularly those obtained from GGA+U and Hybrid functionals.
(3) The feasibility of approximating the fully converged QPscGW results with the GGA+U  calculations for a specific regime of Hubbard U values for simulation of large scale atomic arrangements. 

Finally, we have performed extensive calculations of the fully converged QPscGW calculations with the initial wavefunctions from the GGA+U and Hybrid functional approach. Figure 6 displays the convergence of the QPscGW calculations starting from GGA+U \citep{Kobayashi} with (a) U = 4.0 eV, (b) U = 8.0 eV. A comparison of the converged density of states reveals that the results for GGAU(4.0)+QPscGW \citep{Rinke, Kioupakis} provides the density of states, which agrees significantly with the XPS data \citep{XPS1-Transition-metal-oxides, XPS2-Transition-metal-oxides}. This includes the energy of the peaks, band edges and the amplitude of the peaks \cite{BIS-Transition-metal-oxides} (cumulative average of the peaks provides a measure of the total number of electrons in the system). A comparison of the number of self-consistent iterations required to converge the calculations shows a total of about 40 iterations, and 65 iterations respectively for the GGAU(4.0)+QPscGW, and GGAU(8.0)+QPscGW formalism. Thus, GGA+U with Hubbard parameter U = 4.0 eV provides a better starting wavefunction and hence faster convergence of the quasiparticle GW method. Further, the location of the peaks as well as the band gap obtained from GGAU(4.0)+QPscGW matches well with the experimental XPS results. As mentioned in Figure 3(b), the application of the Hubbard interaction on the Fe d electrons results in the generation of the upper and lower Hubbard bands. As the U value is increased, the band edges are further split resulting in the increase in the predicted band gap. However, we would like to point out that with the increase in the band gap beyond U = 8.0 eV, a band crossing takes place between the Fe $d$ and O $p$ states, resulting in the starting wavefunction which is qualitatively different from that of the previous ones. The authors noted that GGAU calculations with U = 10.0 eV, resulted in converged density of states which were somewhat different qualitatively from the previous calculations with lower Hubbard U values. While the peaks of the occupied bands from calculations match that of the experimental XPS, the amplitudes of the converged QPscGW calculations were significantly higher than the experimental results. 

\begin{table}
\begin{tabular}{|l  c  c  c  c  c  c |} 
\hline\multicolumn{7}{|c|}{ Band gaps and magnetic moments in Iron monoxide} \\

 Theoretical Method& GGA & GGA+U(4) & HSE03 & mGW \citep{Asahi} & Present work (QPscGW) & Experiment \\ 
        
 \hline
 Band gap (eV) & 	0.0 & 1.8 & 2.2 & 2.32 & 3.7 & 2.4 \citep{Bowen} \\ 
 Magnetic moment $(\mu_{B})$ & 3.4 & 3.9 & 3.6 & 3.61 & 3.55 & 3.32 \citep{Roth}	 \\ 
 \hline
\end{tabular}
\caption{The tabular data displays the experimental band gaps and magnetic moments in the transition metal monoxide compared to the computational results obtained from various levels of theoretical approximations including the Kohn-Sham equations, Hubbard interaction, Hybrid functionals and Quasiparticle Green's function approach.
}
\label{Table1}
\end{table}

Interestingly, the energy band gap in the correlated oxide $FeO$ primarily occurs in the minority spin channel $t_{2g}$ manifold of states, whereas, a local approach to the exchange correlation functional was unable to provide a gapped state in this material. Even though, the DFT artificially stabilized by the Hubbard U interaction circumvents the issue \citep{Choice-U2}, the inability of the Kohn Sham approach is inherently due to the semilocal treatment of the exchange correlation functional. This drawback is remedied by incorporating the effects of spatial nonlocality within the so-called Hybrid functionals, resulting in a favourable description of the strongly localized transition metal $3d$ electronic states. The inclusion of nonlocal electron-electron interactions in the exchange correlation functions for the Hybrid functions provide better a better picture of the correlated compound and the local magnetic moments are emhanced compared to that of the Kohn Sham approach \citep{Kohn}. 

Albeit, the fact that the hybrid functional approach provides an accurate choice of wavefunctions for the localized states in $FeO$, it remains to be tested whether it serves as the ideal initial wavefunction for the convergence of the Dyson's equations. Figure 7 displays the convergence of the electronic density of states starting from the two different flavours of the hybrid functional approach \citep{HSE, HSE06} (a) the HSE03 and (b) the HSE06 formalism. A rigorous self-consistent iteration of the electronic states provides convergence of the QPscGW equations with upto 20 steps, and the formalism has been displayed for additional iteration steps upto 40 self-consistent iterations. The hybrid functional+QPscGW method provides us with accurate wavefunctions which match significantly better with that of the experimental XPS results, as well as the position and amplitudes of the valence states. Both the HSE03+QPscGW and HSE06+QPscGW methods accuarately describes the four peaks in the spectra of occupied states and the two higher peak structure in the unoccupied spectra in agreement with the experimental XPS and BIS spectrum of the many-electron wavefunctions respectively. In this connection, it is crucial to mention that there are two flavours of the QPscGW formalism, one where the eigenvalues are updated resulting in accurate quasiparticle energies only, whereas the second one updates the eigenvalues and eigenstates at every iteration. In this study the second formalism was followed, resulting in accurate determination of the peaks as well as the matrix elements thus leading to an accurate prediction of the oscillator strengths for the determination of the optical absorption spectra \citep{Zaanen} in this material. The details of the calculations for the two-particle Green's functions and solution of the Bethe Salpeter equations \citep{BSE, Bechstedt-BSE} for accurate prediction of the optical spectra will be the subject of a later journal submission. 

Finally, we focus on the comparison of the band gaps obtained from the various levels of theoretical approximations presented in Table 1 above. We note that the magnetic moments obtained from first principles and quasiparticle methodologies agree well with the experimental magnetic moments obtained from transport measurements. The fact that the first principles calculations within the GGA yields magnetic moments of 3.4 $\mu_{B}$, which agrees well with the experimental findings, even though the method wrongly predicts the oxide to be metallic. We note that the addition of Hubbard interaction with U = 4.0 eV results in the overestimation of the magnetic moments \citep{David-Singh-2}, compared to that of the Hybrid functionals. Further, the iterative solution of the Dyson's equations implemented within the quasiparticle self-consistent GW methodology (QPscGW) \cite{Hedin, Aryasetiawan} provides magnetic moments similar to those predicted from the hybrid functional calculations. The above observation suggests that application of the Hybrid functionals yield better localization of the correlated electronic states compared to the Kohn Sham equations \citep{Kohn, Hohenberg} artificially stabilized by the Hubbard interaction. We are aware that Density functional theory inaccurately predicts the ground state of the Iron monoxide to be metallic, contrary to the observation of a experimental optical band gap of 2.4 eV. 

\begin{figure*}[ht!]
	\centering 
	\includegraphics[width=0.9\textwidth, angle=0]{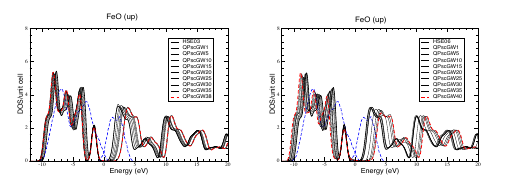}	
	\caption{
The Hybrid functionals (a) HSE03 and (b) HSE06 provide excellent quality of wavefunctions for investigating the electronic
properties near the fermi level for crystals and surfaces. The quick convergence of the density of states with the inclusion of
self-consistent GW corrections
establishes the Hybrid functionals as the best level of theory for the study of Iron oxides and
relevant problems. 
} 
	\label{fig_6}%
\end{figure*}

Measurement of conductivity hints at the band gap to be slightly less than results obtained from optical spectroscopy. Interestingly, very few experimental measurements of the density of electronic states exist in the literature for Iron monoxide. It is challenging to predict the combined experimental data from Photoemission spectroscopy (XPS) and Bremsstrahlung Isochromat Spectroscopy (BIS) \citep{XPS1-Transition-metal-oxides,XPS2-Transition-metal-oxides, BIS-Transition-metal-oxides} to obtain the band gap in this material primarily due to the lack of accuracy in the precise tailoring of the position of the Fermi level from the two different experimental probes. While the optical band gap is obtained to be 2.4 eV, the actual band gap is thought to be slightly less than the measured value. 
The table reflects a comparison of the band gaps obtained from the DFT+U, and HSE03 functionals prioir to self consistently solving the Dyson's equation for obtaining the converged one particle Green's functions. We note that the Hybrid functional approach provides a band gap which is closest to that of the experimental band gap indicating that it provides the best initial wavefunctions for the self consistent GW method. Even though the GGA+U band gap,  approximates the experimental band gap reasonably well, however, the choice of U values are not precisely determined. The Hubbard parameter could be tuned independently, resulting in band gaps which are significantly biased by the choice of the Hubbard U parameter \citep{Choice-U1, Choice-U2}. Further, even though the hybrid functionals provide accurate wavefunctions, and their derivatives, as concluded from the accurate match of the magnetic moments and band gaps from this methodology, the band gaps obtained from fully self consistent GW calculations (HSE03+QPscGW) overestimates the band gap significantly. The reason for this is as follows: The self consistent GW method approximates the self-energy due to electron-electron interactions as a convolution of the quaiparticle Green's function and the screened Coulomb approximation $\Sigma=i\,G W$, thereby leaving out the vertex corrections in the equation $\Sigma=i\,G W \Gamma$ \citep{Schilfgaarde}. Such approximations are valid when the three point vertex renormalization correction is approximated as the product of delta functions $\Gamma(12,3) = \delta(1,2) \;\delta(2,3)$. As a result of neglect of vertex corrections, primarily to avoid complexity leading to reduced computational efficiency, the electron-electron interactions are not scrrened acurately. Such underscreening of the electronic interactions lead to the overestimation of the band gaps. A recent study observed significant decrease in the $GW$ band gap due to the inclusion of vertex correction in prototypical transition metal oxides.

\section{Discussions and Conclusions}
The primary objective of this study is the search for a robust yet efficient computational methodology, which provides the best tradeoff between computational efficiency and accuracy in real materials, particularly that of transition metal oxides. The popular GGA implementation of the Kohn Sham equations erroneously predicted the ground state of Iron monoxide to be metallic, due to the improper treatment of the localized electrons. In order to accurately model the correlated transition metal oxide, several methods such as the DFT artificially stabilized by the Hubbard interaction U \citep{Hubbard} , the range separated Hybrid functionals HSE03 and HSE06 \citep{TunedHSE-Pozun, Moussa-HSE, Marques1}, B3LYP functional \citep{B3LYP,B3LYP1}, and the wavefunction based Hartree Fock method were employed. It was noted that all of these methods remedied the band gap prediction inaccuracies inherent in the Kohn Sham equations. In order to further establish a parameter free theory that accurately incorporates the effect of interactions, we solved the self-consistent Dyson's equation starting from the various levels of theory for correlated electronic systems.

Quasiparticle GW approximation, when applied to the starting wavefunctions from the GGA method resulted in a partial failure to predict the correct wavefunctions and density of states for the correlated electrons. This is primarily due to the incapability of the Hamiltonian to converge to a gapped state starting from a metallic system predicted by the spin resolved GGA  approximation. 
While the GGA+U method predicted accurate wavefunctions and magnetic moments, t therby alleviating the band gap problem, the formalism suffers from the fact that it is heavily biased on the choice of the Hubbard parameter U. More often than not, the Hubbard U values are assigned on an ad-hoc manner, at best guided by experimental results. The bounds on the aplicability of the range of U values are enforced by the correct orbital ordering of the converged electronic states, a criterion which is imposed by the adiabatic evolution of quantum states in first principles calculation of functional materials. Bayesian estimation \citep{Noa} and linear response calculations \citep{Cococcioni, Blugel} further provide estimates of the Hubbard U values which often fails to match the estimated U values for transition metal oxides providing accurate band gaps and magnetic moments. The Hybrid functionals HSE03 and HSE06, \citep{Kresse-HSE} on the contrary were observed to be excellent parameter-free choices of initial wavefunctions for the quasiparticle self consistent GW methodology, providing accurate band edges, band gaps, and spectral fingerprints of the occupied density of states. 

Finally, we have performed a detailed benchmarking of the monotonicity of the convergence of the self consistent GW approximation and the number of iterations required to converge the GW method was treated as a marker to evaluate the proximity of the initial wavefunction to that of the final converged solution. It was observed that the Hybrid functions converged fastest (about 20 iterations), followed by the GGA+U with Hubbard U = 4.0 eV (about 40 iterations) and finally superceded by the B3LYP functional (about 50 iterations). Our first principles calculations amalgamated with the qusiparticle self consistent GW methodology results indicate the two particularl flavours of the Hybrid functional method namely HSE03 and HSE06 as potentially the best tradeoff between computational efficiency and accuracy of wavefunctions. Further, they have successfully provided accurate predictions of the magnetic moments, density of states and band gap in the correlated oxide.  Hence, we suggest the HSE03+QPscGW as the most efficient methodology for accurately describing correlated electron systems, particularly that of transition metal oxides. Inspite of the recent success of the GW methodology, we observed that it is crucial to include the vertex corrections in the self energy for acurate prediction of the band gaps and spectral weights of the density of states. Further, our calculations reveal that it is not beneficial to perform a single shot $G_{0}W_{0}$ \citep{Rodl-G0W0+HSE} calculation starting from the range separated Hybrid functional method, and a fully self-consistent GW approximation needs to be enforced to obtain wavefunctions and their derivatives. This is primarily required to avoid erroneous prediction of the optical oscillator strengths and transition matrix elements \citep{BSE, Rocca-BSE, Fuchs2008} in the transition metal oxide, Iron monoxide and will be elucidated in detail in a later publication.


\section*{Acknowledgements}
The author would like to thank Prof. Efstratios Manousakis and Prof. Emmanouil Kioupakis for valuable discussions regarding the $GW$ methodology. The author acknowledges the High Performance Computing divisions at the NHMFL, Florida State University, and computing cluster at George Mason University for providing access to advanced computing facilities. Finally, the author would like to acknowledge support by BITS Pilani Hyderabad for Institutional facilities, quantum softwares and computational resources.


\newpage
\def\bibsection{\section*{\refname}}
\renewcommand{\bibfont}{\fontsize{10}{10}\selectfont}
\bibliographystyle{elsarticle-num}
\bibliography{algo}

\end{document}